# The origin of the non-monotonic field dependence of the blocking temperature in magnetic nanoparticles

R. K. Zheng,[1] Hongwei Gu,[2] Bing Xu,[2] and X. X. Zhang[1*]

[1]Department of Physics, [2]Department of Chemistry, The Hong Kong University of Science and Technology, Clear Water Bay, Kowloon, Hong Kong, China

The field dependence of the peak temperature of the zero-field-cooled (ZFC) magnetization curve of a magnetic nanoparticle system was studied using a diluted magnetic fluid composed of FePt nanoparticles. It is found that the peak temperature increases with increasing applied field below 3 kOe; it then decreases with further increasing the applied field. The non-monotonic field dependence of the peak temperature in magnetic particle systems is attributed to the anisotropic energy barrier distribution of the particles, and to the slowly decreasing magnetization (or to the non-Curie's law dependence of magnetization) above the blocking temperature. The non-Curie's law dependence of the magnetization is caused by large magnetic anisotropy and Zeeman energy of particles in high magnetic fields. Numerical simulation results, based on basic thermodynamics, and pure thermal relaxation and energy barrier distribution extracted from the low-field, experimental ZFC data show a good agreement with experimental results.

PACS: 75.75.+a; 75.50.Tt; 75.20.–g; 75.50.Ss

---

[*] Email: phxxz@ust.hk

Small magnetic particles have been studied very intensively for a long time due to their fundamental as well as technological relevance [1-3]. With decreasing particle size, the magnetic stability of nanoparticles will become an important issue in technological applications due to thermal agitation [4]. A magnetic nanoparticle is generally in a single domain state with uniaxial anisotropy [2]. Its magnetic moment $\vec{\mu} = \vec{M}_s V$, where $M_s$ and V are the spontaneous magnetization and volume of the particle, either points "up" or "down" in a zero field, if the easy axis is along the Z-axis. The relaxation time for $\vec{\mu}$ between "up" and "down" in a zero magnetic field is determined by the exponential law:

$$\tau = \tau_0 \exp(U/k_B T), \qquad (1)$$

where $\tau_0$ is the attempt frequency of the order of $10^{-9} - 10^{-12}$ s [5], $k_B$ is the Boltzmann constant, and T is the temperature in Kelvin. The anisotropy energy, U=KV, is the energy barrier between the degenerated double-well potential, where K is the anisotropy constant. If the moment is detected by a technique with a measuring time, $t_m$, the particle behaves superparamagnetically when $\tau < t_m$; the particle is in the blocked state when $\tau > t_m$. The blocking temperature, $T_B$, is then defined by $t_m = \tau = \tau_0 \exp(U/k_B T)$, or $T_B = U/[k_B \ln(t_m/\tau_0)]$.

For a sample composed of identical particles with the easy axes aligned in the same direction, U and $\tau_0$ can be extracted from the temperature-dependent ac susceptibility measurements with different frequencies [6]. The blocking temperature or the energy barrier can also be obtained from low field zero-field-cooled (ZFC) and field-cooled (FC) magnetization curves [6,7]. It is worth noting that the magnetic field in the ZFC magnetization measurements should be small enough not to significantly change the energy barrier,

$$U = KV(1 - H/H_K)^2, \qquad (2)$$

where H is the applied magnetic field and $H_k = 2K/M_s$ is the anisotropy field. Because of the size distribution of the particles, there is always an energy barrier distribution in a real particle sample. Even with identical particles, the random orientation of the easy axes will certainly lead to an energy barrier distribution, when a magnetic field is applied to the particle system. In a non-interacting particle sample, the peak temperature, $T_P$, in a zero-field-cooled magnetization curve is simply the average blocking temperature of all the particles. The appearance of the peak temperature can be understood as the following. Particles are first cooled, in a zero field, down to a low temperature, $T_0$ (much lower than $T_P$), at which the net magnetic moment of the sample is zero. When a small field, H, is applied, the sample gains a small net moment due to the fact that the moments of the particles with $T_B \leq T_0$ reach their thermal equilibrium. The particles with larger $T_B$ do not contribute to the net moment in a short time due to the very long relaxation time according to Eq. (1). With increasing T, more and more particles become unblocked and reach their equilibrium, which leads to an increase in the net moment. The increase in the net moment at a particular temperature below $T_P$ is only due to the relaxation of the particles whose blocking temperature is equal to or just slightly higher than the temperature, because the relaxation rate is extremely small for the much larger particles. However, with increasing temperature, the net moment of the particles that are already superparamagnetic (or unblocked) decreases following Curie's law, 1/T. Therefore, with increasing T, a peak must appear in the ZFC. The peak temperature can then be considered as the average blocking temperature of the whole sample. At temperatures well above $T_P$, M(T) should follow Curie's law.

From Eq. (2), it is expected that with an increase in the magnetic field, the energy barrier will monotonically decrease for all particles. Accordingly, their blocking temperatures should shift to lower temperatures. Consequently, the peak in the ZFC curve should also shift to a lower temperature for a larger field. It is straightforward to expect that the peak temperature, $T_P$, decreases monotonically with increasing applied magnetic field.

However, non-monotonic field dependence of the peak temperature was observed in a number of systems, such as ferrimagnetic $Fe_3O_4$ [8] and $\gamma$-$Fe_2O_3$ [9], antiferromagnetic ferritin [10-12], and the molecular magnet of $M_{12}$-ac [13]. The periodic oscillation of the field dependence of $T_P$ in the molecular magnet [13] has been successfully interpreted in terms of resonant spin tunneling [13,14]. This anomaly observed in antiferromagnetic ferritins [10-12] may also be associated with resonant spin tunneling due to the small number of uncompensated spins [12,14]. However, the origin of non-monotonic field dependence of the peak temperature in ferrimagnetic $Fe_3O_4$ [8] and $\gamma$-$Fe_2O_3$ [9] particles remains unclear [14]. In this letter, we present observations of the non-monotonic field dependence of the peak temperature in magnetic FePt nanoparticles and a model to account for the anomaly.

The sample used in this study was the FePt nanoparticles fabricated following the method reported by Sun et al [15,16]. A high-resolution transmission electron microscopy image shows that the nanoparticles have a spherical shape with an average diameter of 2.6 nm and a narrow size distribution. The corresponding diffraction rings have been indexed as the $L1_0$ phase of FePt, as shown in the inset of Fig. 1. The FePt nanoparticles were dispersed in hexane for the magnetic measurements. The volume fraction of the nanoparticles was less than 1% to avoid interparticle interactions. The FePt nanoparticle solution was transferred to a Quantum Design SQUID magnetometer at room-temperature and then a field of 50 kOe was applied to align the easy axes of the nanoparticles in the field direction. After the sample was cooled to 150 K, at which hexane froze into a solid and the particles were in the superparamagnetic state, the 50 kOe field was turned off. The ZFC magnetizations were then measured from 5 K to 150 K in different fields. The representative ZFC curves are shown in Fig. 2(a). The field-dependent peak temperatures, $T_P$, of the ZFC curves can be easily obtained from these curves, as shown in Fig. 3. The intriguing feature of the curve in Fig. 3 is that the peak temperature does have a non-monotonic field dependence as observed in other systems [8-12], i.e., it decreases with the applied field when $H > 3$ kOe and increases with increasing H for $H < 3$ kOe.

As we discussed above the peak in a low-field ZFC curve is due to the competition between the decrease in moments of the superparamagnetic particles (m~ 1/T) and the increase in the net moment due to the newly relaxed larger particles with increasing temperature. It is well known that the 1/T dependence of magnetization is an approximation of the Langevin function and only valid when the Zeeman energy is much smaller than the thermal energy, i.e. $\mu H \ll k_B T$ [17]. In Fig. 2, it is evident that, for large fields, M(T) curves show a much slower decrease than the 1/T dependence even at temperatures well above the peak temperatures. This weak temperature dependence is due to the larger Zeeman energy resulting mainly from the large magnetic moment of the magnetic nanoparticles, $\mu \sim 1000\ \mu_B$, for FePt particles that are 2.6 nm in diameter. Another factor that leads to the non-Curie law may be the strong anisotropy of the magnetic nanoparticles, since the Langevin function is based on the isotropic magnetic moment. As we observed in the $Fe_8$ molecular magnets, the anisotropy plays a very important role in determining the temperature- and field-dependent magnetization curves [18].

It is therefore possible that the non-monotonic field dependence of the peak temperature in magnetic particles (with large moments and anisotropy) could be due to the size distribution and the slowly decreasing magnetization (or invalidity of the Curie law) in high magnetic fields. The low-field ZFC curves have been numerically simulated using the different energy barrier distributions (or size distributions) by taking the Curie's law dependence of moments into account for superparamagnetic particles, which shows qualitative agreement with the experimental curves [9,19]. However, it should be noted that the equations used in those simulations [9,19] did not include the relaxation effect of the moments (or the time-dependent nature of the ZFC magnetization curve). To avoid these problems, we begin with the basic dynamics of the moment to study the non-monotonic field dependence of the peak temperature.

For simplicity, we consider a particle system composed of N, identical and non-interacting magnetic nanoparticles whose easy axes are aligned in the same direction. When H = 0, the net moment of the sample is zero, because the moments of N/2 particles point in one direction and the moments of the rest of the particles point in the opposite direction. When a magnetic field is applied along the easy axes of the particles, one of the degenerated (in the double-well potential) states becomes a real ground state, while the other becomes metastable due to the Zeeman energy, $-\mu H$, as shown in Fig. 4. Consequently, the relaxation times will be different for particles in different wells and given by [1]:

$$\tau^{\pm} = \tau_0 \exp\left[KV\left(1\mp H/H_k\right)^2 / k_B T\right]. \tag{3}$$

The magnetic moment of the system is determined by the difference in the number of particles parallel ($N^+$) or antiparallel ($N^-$) to the applied field.

At a given temperature, the rate of the change in the number of particles is $dN^+ = -N^+ dt/\tau^+ + N^- dt/\tau^-$. The magnetic moment of the system at time t after applying field H can be thus written as [1]:

$$m(t) = NVM_S\left[\frac{\tau^+ - \tau^-}{\tau^+ + \tau^-} - \left(\frac{\tau^+ - \tau^-}{\tau^+ + \tau^-} - \frac{N_0^+}{N_0^-}\right)\exp(-t/\tau)\right], \tag{4}$$

where $N_0^+$ ($N_0^-$) is the initial value of $N^+$ ($N^-$), $\tau$ is the effective relaxation time, and $1/\tau = 1/\tau^+ + 1/\tau^-$.

To demonstrate clearly the peak temperature shift with the field, we consider a very simple case, i.e., a sample composed of nanoparticles of the same material of only two different sizes. The volume of a larger particle is twice that of a smaller one. The ratio of the number of the larger particles to that of the smaller particles is 1:9. Suppose that the intrinsic blocking temperature of the small nanoparticles, $T_B^S$, is 10 K and the blocking temperature for the large nanoparticles, $T_B^L$, will be 20 K. Then, the energy barrier distribution can be written as $f(T_B) = 0.9\delta(T_B - 10) + 0.1\delta(T_B - 20)$. When the system is cooled down, in the zero-field, from the high temperature to $T$, the magnetization of the system is 0, because $N_0^+ = N_0^-$. From Eqs. (3) and (4), we simulated the ZFC curve for $H_1 = 0.01 H_K$ and the temperature sweeping rate as 0.5 K/min, as shown in Fig. 5 (a). The ZFC curves peak at 10 K and 20 K for the smaller and larger nanoparticles, respectively. The ZFC curve of the whole sample shows two peaks corresponding to the two particle sizes. It is found that above the blocking temperature, the M(T) in such a small field follows

Curie's law, i.e. the total M(T) and m(T) for each size of particle show 1/T dependence above their blocking temperatures. Since the total volume of the smaller particles is much larger than that of larger particles, the magnetic moment of the smaller particles should be much larger than that of larger particles at their own blocking temperatures, which is clearly seen in the figure. Due to the fast decease of m(T) of the smaller particles as the temperature increases, the total magnetization at the blocking temperature of large particles is smaller than that at the blocking temperature of the smaller particles. Thus the peak temperature of the whole system should be 10 K, corresponding to the blocking temperature of the small nanoparticles. Similarly, the ZFC curve in a field of $H = 0.1 H_K$ was simulated and is shown in Fig. 5(b). As expected, with an increasing magnetic field, the blocking temperatures shift to lower temperatures, i.e. to 9 K and 17.5 K for the smaller and larger particles, respectively, which is in agreement with Eq. 2. However, the peak temperature of the total magnetization curve for the whole sample should be 17 K, because at which the total magnetization is much higher than that at 10 K. The data shown in Fig. 5 demonstrate clearly that the peak temperature in ZFC curves may increase with field. The physics associated with the observation can be understood as the following. Due to the large Zeeman energy, even for the smaller particles, the magnetization curves no longer follow Curie's law (M ~1/T) and the magnetization decreases quite slowly above $T_B$ (Fig. 5). When the large particles become unblocked, the total magnetization increases sharply and it is significantly higher than the peak due to the deblocking of the smaller particles. This certainly shifts the peak temperature of the M(T) of the sample to a higher temperature, although the blocking temperatures of the individual particles shift to lower temperatures due to the reduction in the energy barrier by the magnetic field.

We have demonstrated, by using a sample with a very simple energy barrier distribution, that the size distribution and the non-Curie law dependent magnetization in a large magnetic field may be the cause of the anomaly in the field-dependent peak temperature of particle systems. To compare the model with experimental data shown in Fig. 3, we used the parameters of the sample, such as the size distribution and anisotropy constant, to simulate the ZFC curves in different fields. Using $U = \ln(t_m/\tau_0)k_B T_B$, $T_P = 11.5$ K in 0.1 kOe ZFC curve and 2.6 nm diameter nanoparticles, we found that $K = 2\times10^7$ erg/cm$^2$, which is close to the previous reported value [15]. The saturation magnetization is about $1\times10^4$ emu/cm$^3$ [20]. It is known that the energy barrier distribution (or $f(T_B)$) can be roughly extracted from low-field ZFC-FC curves [9,19]. The extracted energy barrier distribution is shown in the inset of Fig. 3, which is between the normal and lognormal distributions. We then calculated the ZFC curves for different fields by integrating Eq. (4) over the full distribution:

$$m(t) = \int NVM_S \left[ \frac{\tau^+ - \tau^-}{\tau^+ + \tau^-} - \left( \frac{\tau^+ - \tau^-}{\tau^+ + \tau^-} - \frac{N_0^+}{N_0^-} \right) \exp(-t/\tau) \right] f(V) dV . \qquad (5)$$

In the calculation, the temperature sweep rate, 0.5K/min, used in the experiments, was employed in order to include the relaxation effect. Representatives of the calculated ZFC curves and the field-dependent T$_P$ are shown in Fig. 2(b) and Fig. 3, respectively for comparisons with experimental results. The ZFC curves clearly show that, in the high fields due to the larger Zeeman energy (larger than the thermal energy) and the large magnetic anisotropy energy, the magnetization does not follow Curie's law (or decrease with temperature quite slowly). It is evident in Fig. 3 that the numerical results are in quantitative agreement with the experimental results in low the field regime. The discrepancy in the high-field regime between the two sets of data might be due to the fact that the extracted distribution does not fully reflect the real situation. Another reason could be that, in the numerical simulation, the orientations of the easy axes of the particles are aligned in the field direction, but the easy axes in our sample were not perfectly aligned.

In conclusion, non-monotonic field dependence of peak temperatures, $T_P$, was observed in a FePt nanoparticle system. We have calculated the temperature-dependent magnetization curves in the different magnetic fields by considering the pure thermal relaxation of moments. It was demonstrated that the anomaly is due to the size distribution and the slow decrease of the magnetization (or non- Curie's law dependence of magnetization) above the blocking temperature in high fields. We expect that this effect should be observed in other magnetic nanoparticle systems with proper energy barrier distributions. In fact, we have also observed this phenomenon in Co and $Fe_3O_4$ nanoparticle systems [21].

Acknowledgement: This work was supported by grants from the Research Grants Council of the Hong Kong Special Administration Region, China.

References:

[1] J. L. Dormann, D. Fiorani, and E. Tronc, Advances in Chemical Physics, Vol 98 **98**, 283 (1997).

[2] D. L. Leslie-Pelecky and R. D. Rieke, Chem. Mat. **8**, 1770 (1996).

[3] R. H. Kodama, J. Magn. Magn. Mater. **200**, 359 (1999).

[4] V. Skumryev, S. Stoyanov, Y. Zhang, G. Hadjipanayis, D. Givord, and J. Nogues, Nature **423**, 850 (2003).

[5] T. Jonsson, P. Svedlindh, and M. F. Hansen, Phys. Rev. Lett. **81**, 3976 (1998).

[6] X. X. Zhang, G. H. Wen, G. Xiao, and S. H. Sun, Journal of Magnetism and Magnetic Materials **261**, 21 (2003).

[7] X. X. Zhang, J. M. Hernandez, J. Tejada, and R. F. Ziolo, Physical Review B **54**, 4101 (1996).

[8] W. Luo, S. R. Nagel, T. F. Rosenbaum, and R. E. Rosensweig, Phys. Rev. Lett. **67**, 2721 (1991).

[9] R. Sappey, E. Vincent, N. Hadacek, F. Chaput, J. P. Boilot, and D. Zins, Phys. Rev. B **56**, 14551 (1997).

[10] S. Gider, D. D. Awschalom, T. Douglas, K. Wong, S. Mann, and G. Cain, Journal of Applied Physics **79**, 5324 (1996).

[11] J. R. Friedman, U. Voskoboynik, and M. P. Sarachik, Physical Review B **56**, 10793 (1997).

[12] J. Tejada, X. X. Zhang, E. delBarco, J. M. Hernandez, and E. M. Chudnovsky, Physical Review Letters **79**, 1754 (1997).

[13] J. R. Friedman, M. P. Sarachik, J. Tejada, and R. Ziolo, Physical Review Letters **76**, 3830 (1996).

[14] E. M. Chudnovsky, Journal of Magnetism and Magnetic Materials **185**, L267 (1998).

[15] S. H. Sun, C. B. Murray, D. Weller, L. Folks, and A. Moser, Science **287**, 1989 (2000).

[16] H. W. Gu, R. K. Zheng, X. X. Zhang, and B. Xu, Journal of the American Chemical Society **126**, 5664 (2004).

[17] Robert C. O'Handley, *Modern Magnetic materials: principles and applications*. (John Wiley & Sons, INC., New York, 2000).

[18] X. X. Zhang, H. L. Wei, Z. Q. Zhang, and L. Y. Zhang, Physical Review Letters **8715**, 157203 (2001).

[19] J. C. Denardin, A. L. Brandl, M. Knobel, P. Panissod, A. B. Pakhomov, H. Liu, and X. X. Zhang, Phys. Rev. B **65**, 064422 (2002).

[20] X. W. Wu, C. Liu, L. Li, P. Jones, R. W. Chantrell, and D. Weller, J. Appl. Phys. **95**, 6810 (2004).

[21] R. K. Zheng, Hongwei Gu, Bing Xu, X. X. Zhang, (unpublished).

Figure Captions

Fig. 1 HRTEM image of the FePt nanoparticles. Inset: the diffraction rings of the particles are indexed by the $L1_0$ phase of FePt.

Fig. 2 (a) Experimental ZFC curves measured in various fields. (b) Simulated ZFC curves. Peak temperatures are indicated by arrows.

Fig. 3 Experimental (open circle) and simulated (line) $T_P$ dependences on the field. The inset displays the energy barrier distribution extracted from ZFC-FC curves.

Fig. 4 Landscape of the free energy of a single domain particle with an applied field.

Fig. 5 (a) ZFC curves of $H = 0.01 H_K$ for smaller particles (open squares), larger particles (open circles), and their sum (solid line). (b) ZFC curves of $H = 0.1 H_K$. The peak temperature is indicated for each curve.

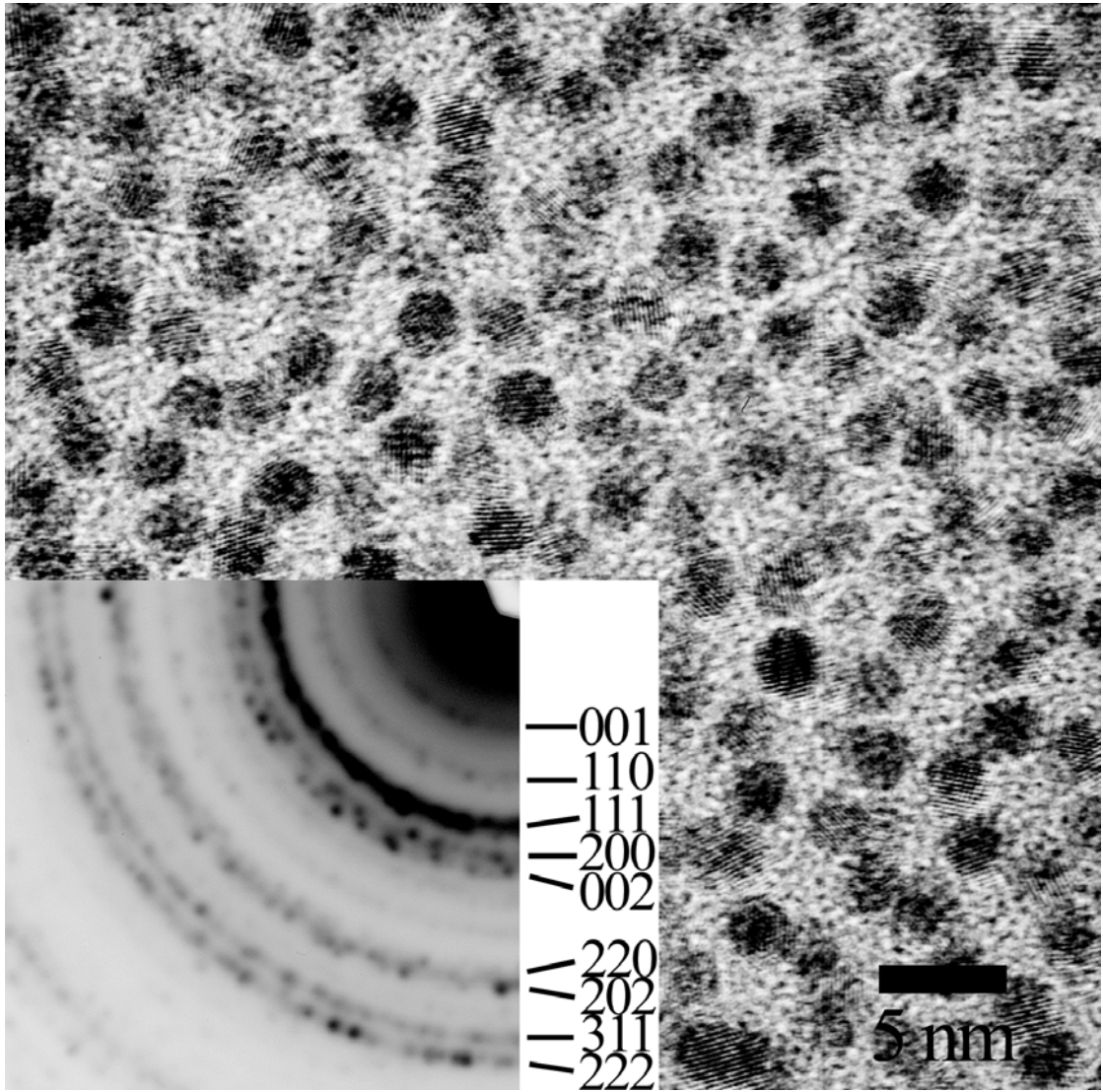

Fig.1

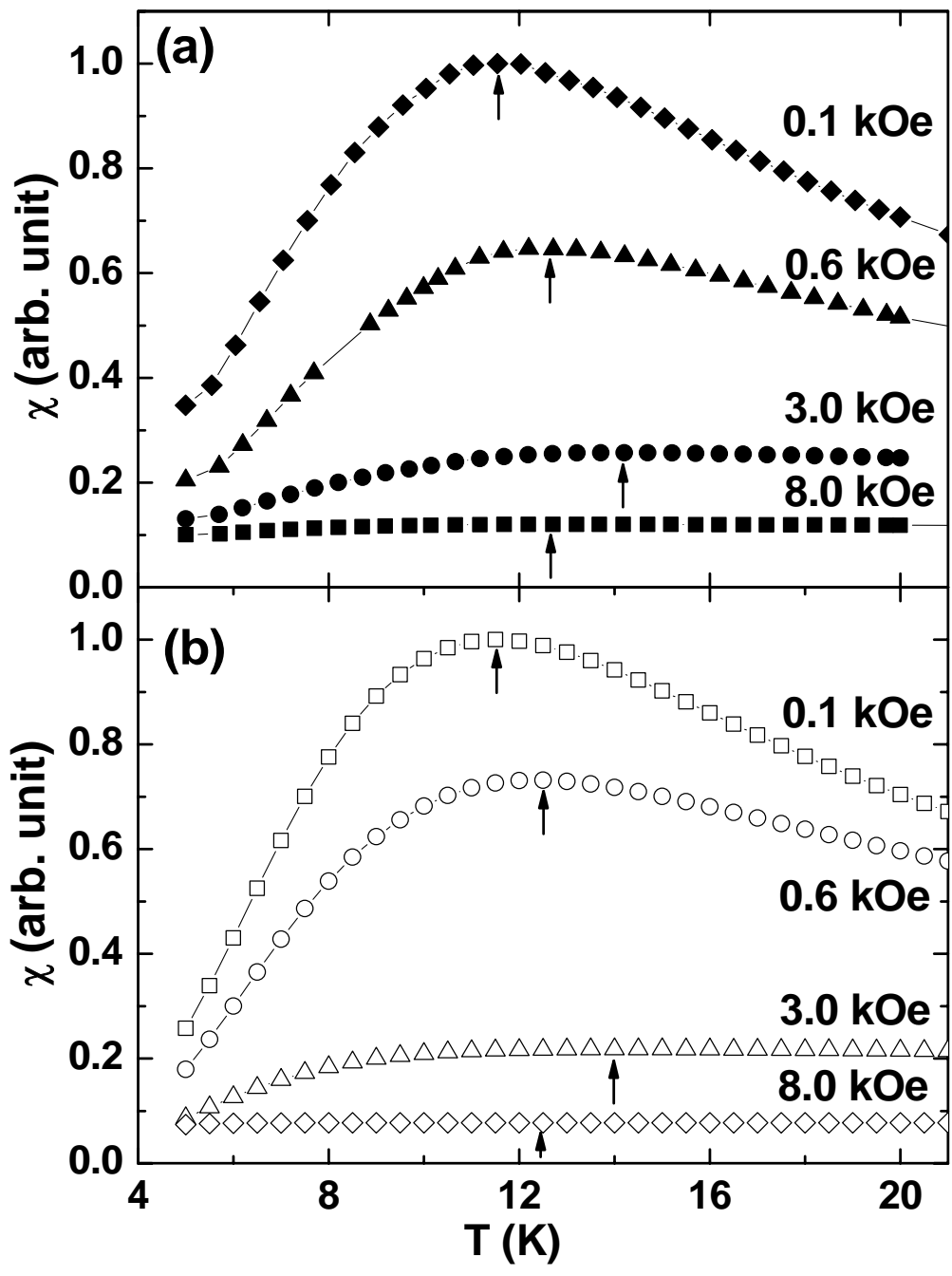

Fig. 2

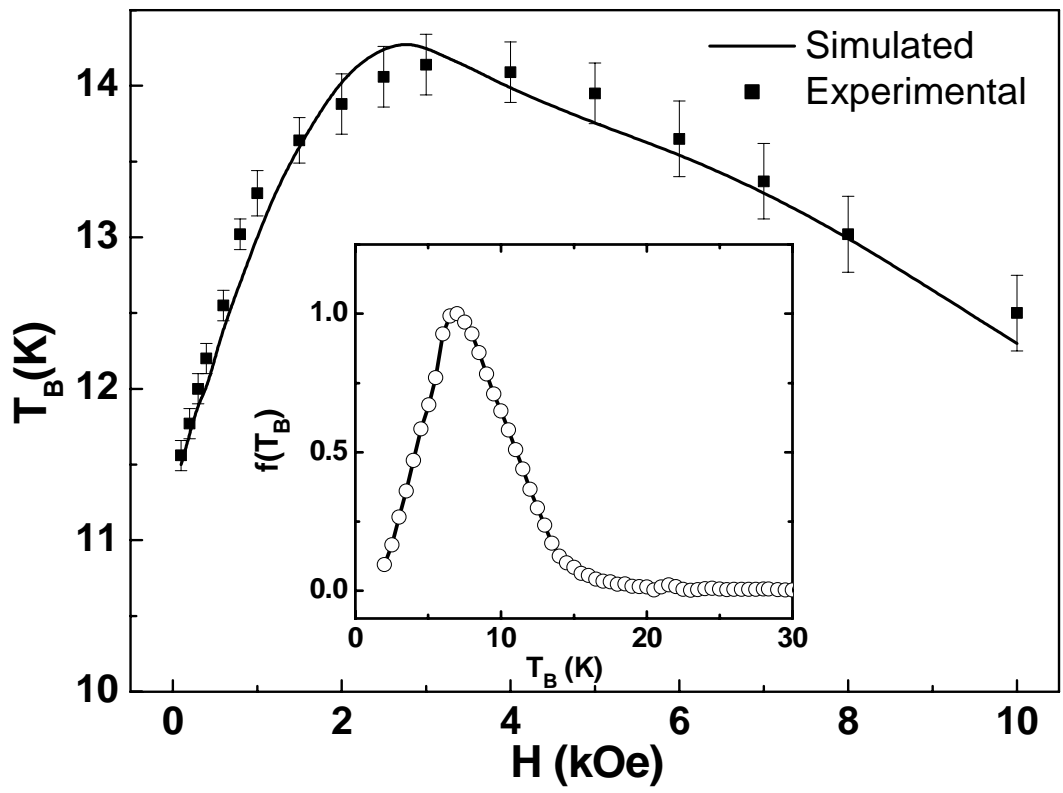

Fig. 3

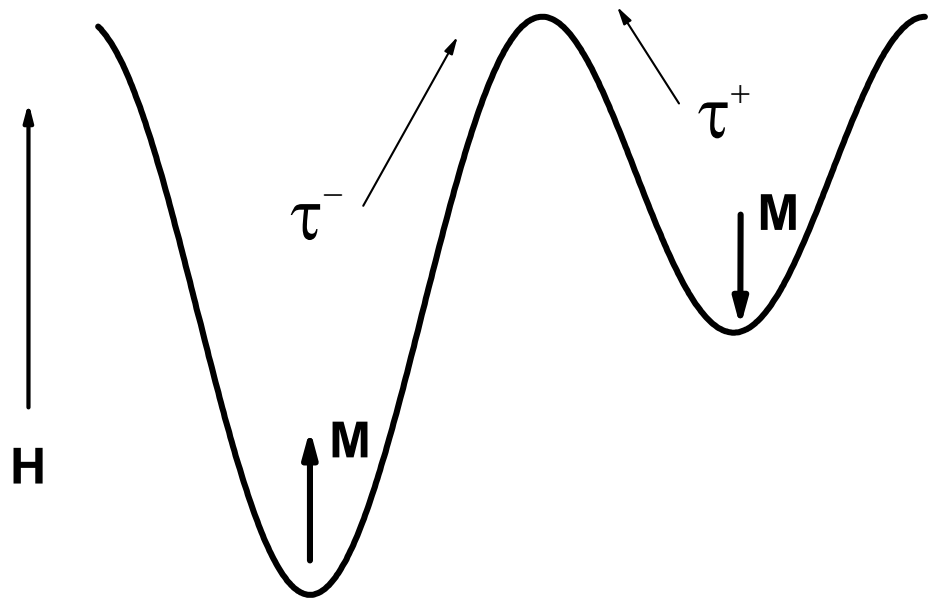

Fig. 4

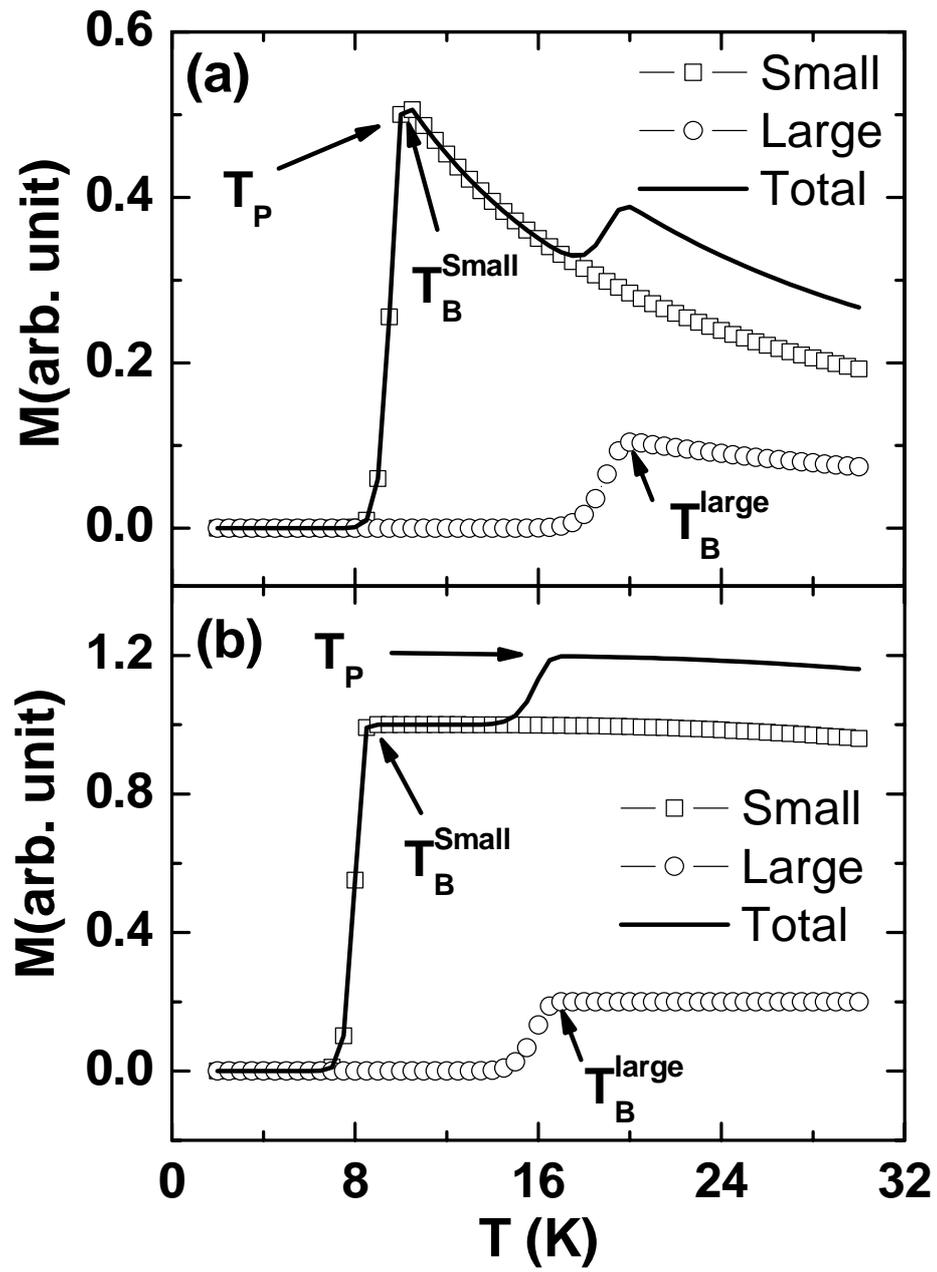

Fig. 5